\documentclass[aps,prl,twocolumn,groupedaddress,showpacs]{revtex4}
\usepackage{graphicx}
\usepackage{wasysym}
\draft

\begin{document}
\
\title{ Quantum Hall Effect in Biased Bilayer Graphene}
\author{R. Ma$^{1,2}$, L. J. Zhu$^2$, L. Sheng$^{3}$,
M. Liu$^{1}$ and D. N. Sheng$^2$}
\address{
$^1$Department of Physics, Southeast University, Nanjing 210096, China \\
$^2$Department of Physics and Astronomy, California State
University, Northridge, California 91330, USA\\
$^3$National Laboratory of Solid State Microstructures and
Department of Physics, Nanjing University, Nanjing 210093, China }

\begin{abstract}
We numerically study the quantum Hall effect in biased bilayer
graphene based on a tight-binding model in the presence of disorder.
Integer quantum Hall plateaus with quantized conductivity
$\sigma_{xy}=\nu e^2/h$ (where $\nu$ is any integer) are observed
around the band center due to the split of the valley degeneracy by
an opposite voltage bias added to the two layers. The central
($n=0$) Dirac Landau level is also split, which leads to a
pronounced $\nu=0$ plateau. This is consistent with the opening of a
sizable gap between the valence and conduction  bands. The exact
spectrum in an open system further reveals that there are no
conducting edge states near zero energy, indicating an insulator
state with zero conductance. Consequently, the resistivity should
diverge at Dirac point. Interestingly, the $\nu=0$ insulating state
can be destroyed by disorder scattering with intermediate strength,
where a metallic region is observed near zero energy.  In the strong
disorder regime, the Hall plateaus with nonzero $\nu$ are destroyed
due to the float-up of extended levels toward the band center and
higher plateaus disappear first.

\end{abstract}

 \maketitle

\section{I. Introduction}

The discovery of an unusual quantum Hall effect (QHE) in bilayer
graphene has stimulated great interest in the study of the
electronic transport properties of this new material~\cite{K. S.
Novoselov,R. V. Gorbachev,S. V. Morozov,E. A. Henriksen,E. McCann,J.
Nilsson,J. G. Checkelsky,Y. Hasegawa,D. A. Abanin,E. V. Gorbar,H.
Min,E. V. Castro,R. Ma}. At low energies and long wavelengths, the electrons in
bilayer graphene can be described in terms of massive, chiral, Dirac particles. While
previous studies have focused on unbiased and thus gapless bilayer
graphene, recent experimental and theoretical studies~\cite{T.
Ohta,Castro,Oostinga,F.Guinea,Min,McCann} have revealed some
interesting aspects of biased bilayer graphene. It has been
shown that an electronic gap between the valence and conduction
bands opens up at the Dirac point and the low energy band acquires a
Mexican hat dispersion relation by changing the density of charge
carriers in the layers through the application of an external field
or by chemical doping, which creates a potential difference between
the layers.  The
presence of the potential bias transforms the bilayer graphene into the
only known semiconductor with a tunable energy gap and may open a
way for developing photodetectors and lasers tunable by the electric
field effect.

Under strong perpendicular magnetic field,  experimental results have
shown that biased bilayer
graphene exhibits a pronounced plateau at zero Hall conductivity
$\sigma _{xy}$=0, which is absent in the unbiased case and can only
be understood as due to the opening of a sizable gap between the
valence and conduction bands~\cite{Castro}. Tight-binding
calculations have shown that the existence of such a gap can have a
significant effect on the Landau level (LL)
spectrum~\cite{McCann, Castro}. While disorder effect is
known to be crucial in the conventional QHE systems, in-depth
understanding of the properties of the QHE in the presence of
disorder in biased bilayer graphene is still absent and hence
greatly needed.

In this work, we carry out a numerical study of the QHE in biased
bilayer graphene in the presence of disorder based upon a
tight-binding model. The Hall conductivity near the band center
exhibits a sequence of plateaus at $\sigma_{xy}=\nu e^2/h$ where
$\nu$ is an integer, as in the conventional QHE systems. The $\nu=0$
plateau is robust with its width proportional to the strength of
bias, which is consistent with the experimental observation. We
further investigate the effect of random disorder on the QHE by
calculating the Thouless number~\cite{Edwards}. Interestingly, at an
intermediate disorder strength, the energy gap around $E_{f}=0$
disappears, which destroys the $\nu=0$ plateau, and the system
undergoes a transition to a metallic state. In the strong-disorder
(or weak-magnetic-field) regime, the QHE plateaus around the band
center can be destroyed due to the float-up of extended levels
toward the band center. The $\nu=\pm 2$ plateaus are the most stable
ones, which disappear last. Furthermore, we have also calculated the
energy spectrum for an open system (cylindric geometry), and
performed numerically a Laughlin's gauge experiment~\cite{Laughlin,
Halperin} by adiabatically inserting flux quantum to directly probe
the quantum transport near the sample edges. No conducting edge
states are observed in the $\nu=0$ energy gap, suggesting an
insulating state with divergent resistivity.

The paper is organized as follows. In Sec.\ II, we introduce the
model Hamiltonian and formulas for the calculation. In Sec.\ III,
numerical results based on exact diagonalization and transport
calculations are presented. Sec.\ IV concludes with a summary.

\section{II. The tight-binding model of biased bilayer graphene}

We consider a bilayer graphene sample consisting of two coupled
hexagonal lattices including inequivalent sublattices $A$, $B$ on
the bottom layer and $\widetilde{A}$, $\widetilde{B}$ on the top
layer. The two layers are arranged in the AB (Bernal)
stacking~\cite{S. B. Trickey,K. Yoshizawa}, where $B$ atoms are
located directly below $\widetilde{A}$ atoms, and $A$ atoms are the
centers of the hexagons in the other layer. Here, the in-plane
nearest-neighbor hopping integral between $A$ and $B$ atoms or
between $\widetilde{A}$ and $\widetilde{B}$ atoms is denoted by
$\gamma_{AB} =\gamma_{\widetilde{A}\widetilde{B}}=\gamma_{0}$. For
the interlayer coupling, we take into account the largest hopping
integral between a $B$ atom and the nearest $\widetilde{A}$ atom
$\gamma_{\widetilde{A}B}=\gamma_{1}$, and the smaller hopping
integral between an $A$ atom and three nearest $\widetilde{B}$ atoms
$\gamma_{A\widetilde{B}}=\gamma_{3}$. The values of these hopping
integrals are taken to be $\gamma_{0}=3.16$ eV, $\gamma_{1}=0.39$
eV, and $\gamma_{3}=0.315$ eV, the same as in Ref.~\cite{R. Ma}.

We assume that each monolayer graphene has totally $L_{y}$ zigzag
chains with $L_{x}$ atomic sites on each chain~\cite{Sheng}. The
size of the sample will be denoted as $N=L_{x}\times L_{y}\times
L_{z}$, where $L_{z}=2$ is the number of graphene monolayers stacked
along the $z$ direction.
In the presence of an applied magnetic
field perpendicular to the plane of the biased bilayer graphene, the
lattice model in real space can be written the following
form~\cite{R. Ma}:
\begin{eqnarray}
H&=&-\gamma_{0}(\sum\limits_{\langle
ij\rangle}e^{ia_{ij}}c_{i}^{\dagger }c_{j}+\sum\limits_{\langle
ij\rangle}e^{ia_{ij}}\widetilde{c}_{i}^{\dagger}\widetilde{c}_{j})\nonumber\\
&-&\gamma_{1}\sum\limits_{\langle
ij\rangle_1}e^{ia_{ij}}c_{jB}^{\dagger}\widetilde{c}_{i\widetilde{A}}
-\gamma_{3}\sum\limits_{\langle
ij\rangle_3}e^{ia_{ij}}c_{iA}^{\dagger}\widetilde{c}_{j\widetilde{B}} +h.c.\nonumber\\
&+&\sum\limits_{i}(w_{i}+\epsilon_1)c_{i}^{\dagger}c_{i}+(w_i+\epsilon_2)\widetilde{c}_{i}
^{\dagger}\widetilde{c}_{i},
\end{eqnarray}
where $c_{i}^{\dagger}$($c_{iA}^{\dagger}$),
$c_{j}^{\dagger}$($c_{jB}^{\dagger}$) are creating operators on $A$
and $B$ sublattices in the bottom layer, and
$\widetilde{c}_{i}^{\dagger}$($\widetilde{c}_{i\widetilde{A}}^{\dagger}$),
$\widetilde{c}_{j}^{\dagger}$($\widetilde{c}_{j\widetilde{B}}^{\dagger}$)
are creating operators on $\widetilde{A}$ and $\widetilde{B}$
sublattices in the top layer. The sum $\sum_{\langle ij\rangle}$
denotes the intralayer nearest-neighbor hopping in both layers,
$\sum_{\langle ij\rangle_1}$ stands for interlayer hopping between
the $B$ sublattice in the bottom layer and the $\widetilde{A}$
sublattice in the top layer, and $\sum_{\langle ij\rangle_3}$ stands
for the interlayer hopping between the $A$ sublattice in the bottom
layer and the $\widetilde{B}$ sublattice in the top layer, as
described above. For the biased system the two layers gain different
electrostatic potentials, and the corresponding energy difference is
given by $\Delta _g=\epsilon_2-\epsilon_1$ where
$\epsilon_1=-\frac{1}{2}\Delta_g$, and
$\epsilon_2=\frac{1}{2}\Delta_g$. For illustrative purpose, a
relatively large asymmetric potential $\Delta_g=0.05\gamma_0$ is
assumed. $w_{i}$ is a random disorder potential uniformly
distributed in the interval $w_{i}\in \lbrack -W/2,W/2]\gamma_0$.
The magnetic flux per hexagon $\phi =\sum_{{\small
{\mbox{\hexagon}}}}a_{ij}=\frac{2\pi }{M}$ is proportional to the
strength of the applied magnetic field $B$, where $M$ is assumed to
be an integer.

\section{III. Results and Discussion}

The Hall conductivity $\sigma _{xy}$ can be calculated by using the
Kubo formula through exact diagonalization of the system
Hamiltonian~\cite{R. Ma}. In Fig.\ 1, the Hall conductivity $\sigma
_{xy} $ near the band center is plotted as a function of electron
Fermi energy $E_{f}$ for a clean sample ($W=0$) of size $N=96\times
24\times 2$ with magnetic flux $\phi =\frac{2\pi }{48}$, for biased
and unbiased cases. Since the Hall conductivity is antisymmetric
about zero energy, we show it mainly in the negative energy region.
As we can see, in the unbiased case, the Hall conductivity exhibits
a sequence of plateaus at $\sigma _{xy}=\nu e^2/h$, where $\nu
=kg_{s}$ with $k$ an integer and $g_{s}=2$ due to double-valley
degeneracy~\cite{E. McCann,Sheng} (the spin degeneracy will
contribute an additional factor $2$, which is omitted here). The
transition from the $\nu =-2$ plateau to $\nu=2$ plateau is
continuous without a $\nu=0$ plateau appearing in between, so that a
step of height $4e^2/h$ occurs at the neutrality point. However,
when a bias is applied, the valley degeneracy is lifted due to the
different projection natures in the two layers of the LL states in
the $K$ and $K'$ valleys. The valley asymmetry has a strong effect
on the LLs near zero energy, where the charge imbalance is
saturated. As a consequence, the Hall conductivity is quantized as
$\sigma _{xy}=\nu e^2/h$, where $\nu =kg_{s}$ with $k$ an integer
and $g_{s}=1$ for each LL due to the split of double-valley
degeneracy~\cite{McCann}. With each additional LL being occupied,
the total Hall conductivity is increased by $e^{2}/h$. Around the
particle-hole symmetric point $E_{f}=0$, a pronounced plateau with
$\sigma _{xy}=0$ is found, which can only be understood as due to
the opening of sizable gap, $\Delta _g$, between the valence and
conductance bands. The emerged zero Hall plateau is accompanied by a
huge peak in the longitudinal resistivity $\rho_{xx}$, indicating an
insulating state. This behavior has been observed
experimentally~\cite{Castro}. It implies that a diverging
$\rho_{xx}$ at the particle-hole symmetric point $E_{f}=0$, in
striking contrast to all the other Hall plateaus, where $\rho_{xx}$
vanishes as same as in ordinary QHE.

\begin{figure}[tbp]
\includegraphics[width=3.3in]{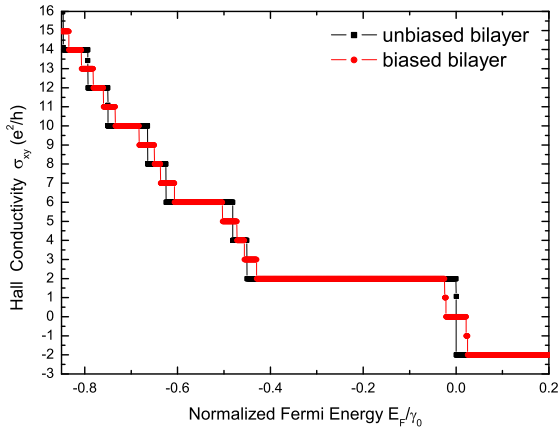}
\caption{Hall conductivity near the band center of unbiased and
biased bilayer graphene with $\protect\phi=\frac{2\protect\pi}{48}$ .
The disorder strength and sample size are set to $W=0$ and
$N=96\times 24\times 2$. Here, the spin degree of freedom has been
omitted.}
\end{figure}

Now we study the effect of random disorder on the QHE around the
band center in the biased bilayer graphene based upon the
calculation of the Thouless number. In Fig.\ 2, the Hall
conductivity $\sigma _{xy}$ and Thouless number $g$ around the band
center are shown as functions of $E_{f}$ for three different
disorder strengths and a relatively weak magnetic flux $\phi
=\frac{2\pi }{48}$. In Fig.\ 2a, the calculated $\sigma _{xy}$ and
Thouless number $g$ at a weak-disorder strength $W=0.2$ are plotted.
Clearly, each valley in the Thouless number corresponds to a Hall
plateau and each peak corresponds to a critical point between two
neighboring Hall plateaus. We will call the central valley at
$E_{f}=0$ the $\nu=0$ valley, the first one just above (below) it
the $\nu=-1$ ($\nu=1$) valley, the second one the $\nu=-2$ ($\nu=2$)
valley, and so on, as same as the Hall plateaus. In Fig.\ 2b, the
Hall conductivity $\sigma_{xy}$ and Thouless number $g$ for a
relatively strong-disorder strength $W=0.6$ are plotted. We see that
the plateaus with $\pm 2$, $\pm 6$ and $\pm 10$ remain well
quantized, and the other plateaus become indiscernible, because of
their relatively small plateau widths. With increasing $W$, higher
valleys in the Thouless number $g$ (with larger $|\nu |$) are
destroyed first, indicating the destruction of the corresponding
higher Hall plateau states. When $W=2.0$, all the plateaus except
for the $\nu =\pm 2$ ones are destroyed (see Fig.\ 2c). The last two
plateaus $\nu =\pm 2$ eventually disappear around $W\sim 3.2$. Thus
we observed that the destruction of the QHE states near the band
center are due to the float-up of extended levels toward zero
energy.

In Fig.\ 3a, we show the Hall conductivity $\sigma _{xy}$ as a
function of $E_{f}$ for a relatively strong magnetic flux $\phi
=\frac{2\pi }{12}$ and three different system sizes $N=24\times
12\times 2$, $N=48\times 24\times 2$, $N=96\times 24\times 2$ at
disorder strength $W=2.0$. We can see that at this disorder strength,
the transition from $\nu
=-2$ plateau to $\nu=2$ plateau becomes continuous. With increasing the
system size, the width of the plateau $\nu=\pm 2$ remains nearly
unchanged. The region around the zero energy of Fig.\ 3a is enlarged
in Fig.\ 3b. For comparison, we also show the results for the
unbiased case, which clearly demonstrate the continuous behavior
between the $\nu =-2$ plateau to the $\nu=2$ plateau in both cases.
This behavior indicates a metallic state occurs around zero energy, which
is essentially caused by the strong coupling between the two Dirac LLs
due to disorder scattering.

\begin{figure}[tbp]
\par
\includegraphics[width=3.4in,height=3.6in]{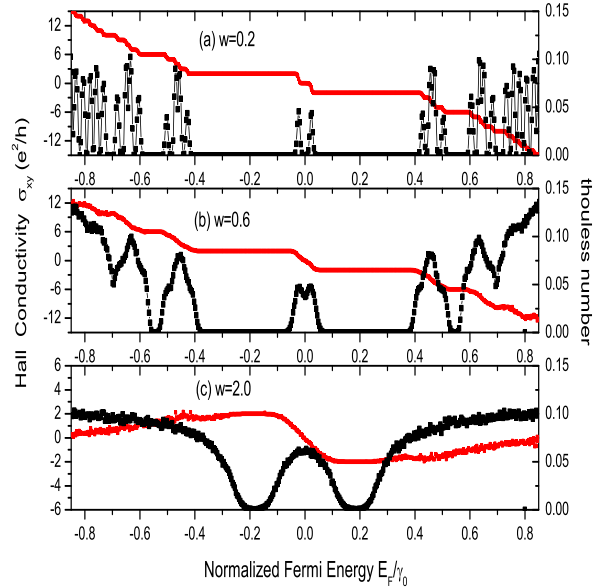}
\caption{Calculated Thouless number and Hall conductivity for
$\protect\phi =\frac{2\protect\pi }{48}$ and three different
disorder strengths, which are averaged over 400 disorder
configurations. Here, the sample sizes are taken to be $N=96\times
48\times 2$ and $N=96\times 24\times 2$ in the calculations of the
Thouless number and the Hall conductivity, respectively.}
\end{figure}

\begin{figure}[tbp]
\includegraphics[width=3.3in]{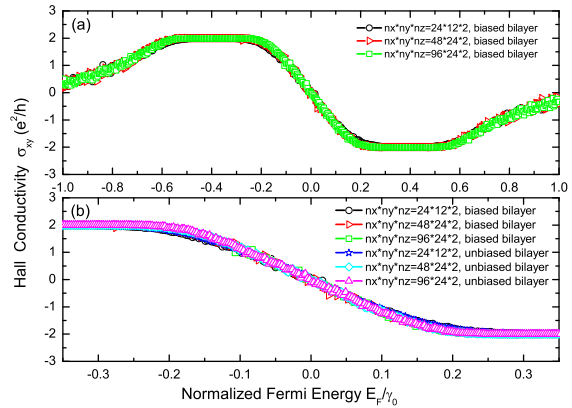}
\caption{Calculated Hall conductivity for biased and unbiased
bilayer graphene with magnetic flux $\protect\phi =\frac{2\protect\pi }{12}$
at disorder strength $W=2.0$ for three different system sizes.}
\end{figure}

We now investigate the evolution of the edge states in an infinitesimal
electric field by performing the Laughlin's gauge
experiment~\cite{Laughlin, Halperin}. A periodic
boundary condition in the $y$ direction and an open boundary
condition in the $x$ direction are imposed to the system.
The system can thus be considered as a cylinder.
When the flux $\protect\theta_y(t)$ threading the cylinder is
adiabatically turned on from $\protect\theta_y(0)=0$ to
$\protect\theta_y(t)$=2$\pi$, which is equivalent to applying a weak
electric field along the $y$ direction
\[
E_y(t)=-\frac {1}{L_y}\frac {\partial\protect\theta(t)}{\partial t }
\].
By diagonalizing the Hamiltonian Eq.(1) under the open boundary
condition along the $x$-axis, at 200 different $\theta_y$, the
eigenenergies $E_n$ of the system are obtained. Fig.\ 4a shows the
calculated energy spectrum as a function of $\protect\theta_y$ for a
clean sample ($W=0$) at system size $N=96\times 24\times 2$. Note
that $\protect\theta_y=0$ and $\protect\theta_y$=2$\pi$ are equivalent, as
the system hamiltonian is periodic
H($\protect\theta_y=0$)=H($\protect\theta_y$=2$\pi$). We first examine the
energy spectrum corresponding to the $\nu=-2$ QHE plateau.
We observe that with changing $\theta_y$, the energy
levels in the plateau region cross each other, which correspond to
two conducting edge channels in accordance with the quantized Hall
effect.
For example, we choose Fermi energy $E_{f}=0.1\gamma_0$. For
$\theta_y=0$, in the ground state, all the single particle states
below $E_{f}$ are occupied, whereas unoccupied above $E_{f}$. Upon
insertion of the flux quantum, the two occupied states below $E_{f}$
are pumped onto states above $E_{f}$ indicated by the arrow, which
causes two electrons transferred across from one edge to the other,
corresponding to the quantized Hall conductivity with $\sigma
_{xy}=2e^{2}/h$, as shown in Fig.\ 4b.  However, there are no such
conducting edge states near $E_{f}=0.0$, where the $\nu=0$ plateau is found.
Clearly, a true spectrum gap shows up corresponding to a trivial
insulating phase, which results in zero net charge transfer, and the
current carried around the ribbon loop is zero.

Now we consider the disorder effect. Fig.\ 5a shows the results for
a randomly chosen disorder configuration for $W=2.0$ at system size
$N=96\times 24\times 2$. We can see that the energy gap around
$E_{f}=0$ disappears. This behavior indicates that the
transition from $\nu =-2$ plateau to $\nu=2$ plateau becomes
continuous, as shown in Fig.\ 5b.  In contrast, if we choose an arbitrary Fermi
energy in the $\nu=\pm 2$ plateau regions, e.g.,
$E_{f}=0.16\gamma_0$, there are always two electrons transferred
across from one edge to the other. Before the $\nu=2$ plateau is
destroyed by the disorder, the $E_f=0$ point becomes metallic.

\begin{figure}[tbp]
\par
\includegraphics[width=3.4in]{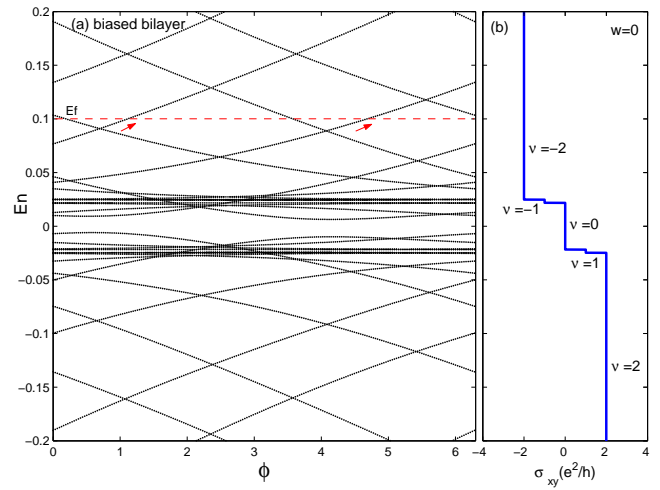}
\caption{(a) Energy levels of biased bilayer graphene with an
open boundary in the $x$ direction, as a function of the twisted boundary phase
$\theta_y$ in the $y$ direction. (b) Hall conductivity near the band
center for $W=0$. Here $\protect\phi =\frac{2\protect\pi }{48}$ and
$N=96\times 24\times 2$.}
\end{figure}

\begin{figure}[tbp]
\par
\includegraphics[width=3.4in]{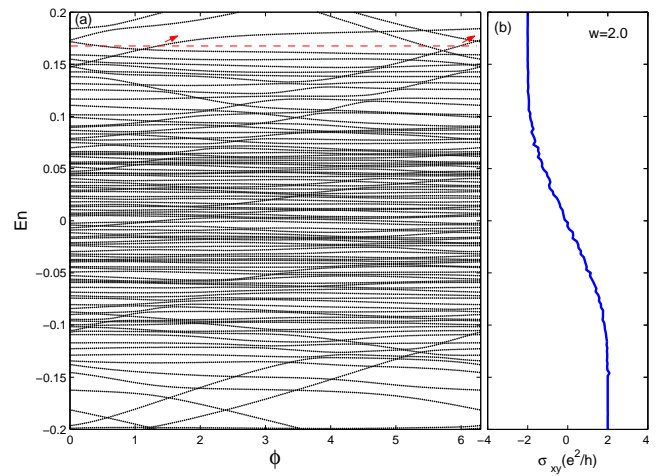}
\caption{(a)Energy levels of unbiased bilayer graphene
as a function of twisted phase
$\theta_y$. (b) (b) Hall conductivity near the band
center for $W=0$. Here $\protect\phi =\frac{2\protect\pi }{48}$ and
$N=96\times 24\times 2$.}
\end{figure}

\section{IV. Summary}

In summary, we have numerically investigated the QHE in biased
bilayer graphene based on tight-binding model in the presence of
disorder. The experimentally observed unconventional QHE is
reproduced near the band center, where the Hall conductivity is
quantized as $\sigma _{xy}=\nu e^2/h$ with $\nu$ being any integer,
including $\nu=0$. The $\nu=0$ plateau around $E_{f}=0$ is due to
the opening of sizable gap between the valence and conductance
bands, which is absent in the unbiased case. By performing
numerically a laughlin's gauge experiment, we have found that there
are no conducting edge states in the $\nu=0$ plateau region, in
contrast to the $\nu\neq 0$ plateaus, where energy levels across
each other, resulting in charge transfer between the edges and
charge accumulation at the edges. However, at an intermediate
disorder strength, the energy gap around $E_{f}=0$ disappears, which
indicates that the transition from $\nu =-2$ plateau to $\nu=2$
plateau becomes continuous, in agreement with the calculated results
of the Hall conductivity. Furthermore, we show that with increasing
disorder strength, the Hall plateaus can be destroyed through the
float-up of extended levels toward the band center and higher
plateaus disappear first. At a strong-critical-disorder strength
$W=W_{c}=3.2$, the most stable QHE states with $\nu=\pm 2$
eventually disappear, which indicates a transition of all the QHE
phases into an insulating phase.

\textbf{Acknowledgment:} This work is supported by the US DOE grant
DE-FG02-06ER46305 (LJZ, DNS), the NSF grant DMR-0605696 (RM, DNS).
We thank the KITP for partial support through the NSF grant
PHY05-51164. We also thank the partial support from the State
Scholarship Fund from the China Scholarship Council, the Scientific
Research Foundation of Graduate School of Southeast University of
China (RM), the doctoral foundation of Chinese Universities under
grant No. 20060286044(ML), the National Basic Research Program of
China under grant Nos.: 2007CB925104 and 2009CB929504 (LS), and the
NSF of China grant No.: 10874066 (LS).

\end{document}